\documentclass[conference]{IEEEtran}
\IEEEoverridecommandlockouts
\pdfoutput=1 
\usepackage[utf8]{inputenc}
\usepackage[T1]{fontenc}
\usepackage{amsmath,amssymb,amsfonts}
\usepackage{algorithmic}
\usepackage{graphicx}
\usepackage{textcomp}
\usepackage[table,xcdraw]{xcolor}
\usepackage{xcolor}
\usepackage{multirow}
\usepackage{tabularx,ragged2e,booktabs,caption}
\usepackage{biblatex} %Imports biblatex package
\usepackage{enumitem}
\usepackage{pifont}
\def\BibTeX{{\rm B\kern-.05em{\sc i\kern-.025em b}\kern-.08em
    T\kern-.1667em\lower.7ex\hbox{E}\kern-.125emX}}
\usepackage{xspace}
\usepackage{flushend}
\usepackage{algorithm2e}
\begin{document}
\RestyleAlgo{ruled}
\title{Variational Quantum Pulse Learning\\
\vspace{5pt}

\vspace{-12pt}
}

\newcommand{\HW}[1]{\textcolor{blue}{[HW: #1]}}

\author{\IEEEauthorblockN{
Zhiding Liang*\textsuperscript{1}  \ \
Hanrui Wang*\textsuperscript{2}  \ \
Jinglei Cheng\textsuperscript{3}  \ \
Yongshan Ding\textsuperscript{4}  \ \
Hang Ren\textsuperscript{5}  \ \
Zhengqi Gao\textsuperscript{2}\\
Zhirui Hu\textsuperscript{6}    \ \
Duane S. Boning\textsuperscript{2}  \ \
Xuehai Qian\textsuperscript{3}  \ \ 
Song Han\textsuperscript{2}  \ \
Weiwen Jiang\textsuperscript{6}  \ \
Yiyu Shi\textsuperscript{1}}
\IEEEauthorblockA{
\textsuperscript{1}University of Notre Dame, IN, USA.
\textsuperscript{2}Massachusetts Institute of Technology,
MA, USA.\\
\textsuperscript{3}University of Southern California, CA, USA.
\textsuperscript{4} Yale University, CT, USA.\\
\textsuperscript{5}University of California, Berkeley, CA, USA. 
\textsuperscript{6}George Mason University, VA, USA.\\
*These authors contributed to the work equally and should be regarded as co-first authors.\\
Corresponding authors: zliang5@nd.edu, hanrui@mit.edu 
\vspace{-0.15in}}
}

\maketitle

\begin{abstract}
Quantum computing is among the most promising emerging techniques to solve problems that are computationally intractable on classical hardware. 
A large body of existing works focus on using variational quantum algorithms on the gate level for machine learning tasks, such as the variational quantum circuit (VQC). However, VQC has limited flexibility and expressibility due to limited number of parameters, e.g. only one parameter can be trained in one rotation gate.
On the other hand, we observe that quantum pulses are lower than quantum gates 
in the stack of quantum computing and 
offers more control parameters. 
Inspired by the promising performance of VQC, in this paper we propose variational quantum pulses (VQP), a novel paradigm to \textit{directly train quantum pulses} for learning tasks. The proposed method manipulates variational quantum pulses by pulling and pushing the amplitudes of pulses in an optimization framework. Similar to variational quantum algorithms, our framework to train pulses maintains the robustness to noise on Noisy Intermediate-Scale Quantum (NISQ) computers. In an example task of binary classification, VQP learning achieves up to 11\% and 9\% higher accuracy compared with VQC learning 
on the qiskit pulse simulators (with system model from real machine) and ibmq-jarkata, respectively, demonstrating its effectiveness and feasibility. Stability for VQP to obtain reliable results has also been verified in the presence of noise.

\end{abstract}


\section{Introduction}
\label{sec1}
Quantum computing has the properties of superposition and entanglement, which grants quantum speedup for certain problems \cite{alexeev2021quantum, ding2020quantum, gokhale2020optimization, niu2020hardware}. Compared with classical computing, exponential speedup has been demonstrated in areas such as quantum chemistry \cite{peruzzo2014variational,cao2019quantum}, finance \cite{ghosh2018identifying}, and machine learning \cite{jiang2021co, wang2021exploration, jiang2021machine}. Therefore, quantum computing has been considered among the best candidates for solving some complex computational problems that cannot be solved by classical computers. In recent decades, technologies such as superconducting transmon that support quantum computers have evolved substantially. 
Google released Sycamore \cite{arute2019quantum} in 2019 and claimed to achieve quantum supremacy, which sampled a circuit with 53 quantum bits (qubits), and the distribution is difficult for a classical computer to simulate in principle. IBM also released a 127-qubit quantum computer at the end of 2021. It is probable that quantum computers with around 1,000 qubits can be manufactured in the coming decade.

The workflow to run programs on quantum hardware can be divided into two subsequent steps. Firstly, the quantum programs are synthesized and compiled into quantum gates. When the programs are sent to the real quantum hardware, the quantum gates will be transformed into quantum pulses with a look-up table. For currently available superconducting quantum computers, the final control signals are in the form of microwave pulses.
A major focus of quantum programs today is the variational quantum algorithm, especially for machine learning tasks, such as RobustQNN \cite{wang2021roqnn} and QuantumFlow \cite{jiang2021co}, which 
uses a classical optimizer to train a parametrized quantum circuit.

Most of the variational quantum algorithms manipulate the parameters at the gate level. For example, quantum neural networks (QNN) encode the input data \cite{wang2021quantumnas, tacchino2020variational, wang2022chip} and perform machine learning tasks on a quantum computer by building and training parametric quantum gate networks. Some state-of-the-art QNNs obtain high accuracy on several classical datasets to demonstrate potential advantage \cite{wang2021quantumnas}. Robust QNNs have also been proposed to mitigate the high noise in NISQ quantum computer \cite{liang2021can}. Decent accuracy has been demonstrated even in presence of noise \cite{wang2021roqnn}. Some recent work tackles the quantum problem with machine learning. For example, reinforcement learning is adopted to find optimal parameters at the circuit level \cite{ostaszewski2021reinforcement}. 

However, variational quantum gates have limited flexibility. A controlled rotation gate comes with only one parameter that can be trained. As pointed out by \cite{sim2019expressibility}, the expressibility and entangling capability of parameterized quantum circuits are mainly affected by the number of parameters. Pulses-level operations, being lower than gate level in the stack of quantum computing, can provide finer  controls and thus granting more flexible manipulation of internal parameters. Therefore, we contemplate that the increased parameters at the pulse level will give us advantages during the training process while maintaining the circuit latency. 
We term such a novel paradigm as {\em variational quantum pulse (VQP) learning}, which though intuitively appealing is in the uncharted territory.   
Note that  a seemingly relevant study is on algorithms to generate pulses to manipulate the qubits \cite{leung2017speedup, khaneja2005optimal, caneva2011chopped, peng2021deep, porotti2022gradient, sivak2021model}. Our goal is fundamentally different. We are not interested in the effective ways to generate pulses, but rather making various 
pulse parameters learnable for machine learning tasks. 

The architecture of VQP learning can be divided into three parts. The first part is to split the traditional QNN into the encoding circuit and trainable circuit (VQC). The encoding circuit is converted into pulse schedule and saved for later use. Then, the trainable circuit is also converted into pulse schedule. But the amplitudes are extracted from the pulse schedule. These amplitudes are the parameters that will be updated in the training process. The second part is to use an optimization framework to iteratively update the amplitudes with the target to minimize the error of classification tasks. The third part is to reconstruct the pulses with group of amplitudes obtained from optimization. Then, the trained VQP can be used for inference tasks. The major contributions of this work can be described as follow:

\begin{itemize}
    %%\item This is the first paper to train pulse parameters in quantum neural networks
    \item This is the first work that verifies the trainability of VQP for machine learning tasks. We train VQP for binary classification tasks and obtain stable results in noisy environments.
    \item Comparing the results of VQC learning and VQP learning for the same tasks demonstrates the advantages of VQP and sheds light on why VQP learning are more promising.
\end{itemize}

The paper is organized as follows. Section \ref{sec1} briefly introduces the background for VQP. Section \ref{sec2} introduces the motivation of this work, gives a definition of VQP, and discusses the possible advantages of VQP learning. Section \ref{sec3} proposes a framework for VQP learning, describes the components of the framework and the optimization methods. Section \ref{sec4} unfolds the experimental results that validate the idea of VQP. Section \ref{sec5} discusses the purpose of extracting pulse amplitudes from the pulse, Section \ref{sec6} concludes the paper with a perspective of the future of VQP learning.

\begin{figure}[t]
\centerline{\includegraphics[width=\linewidth]{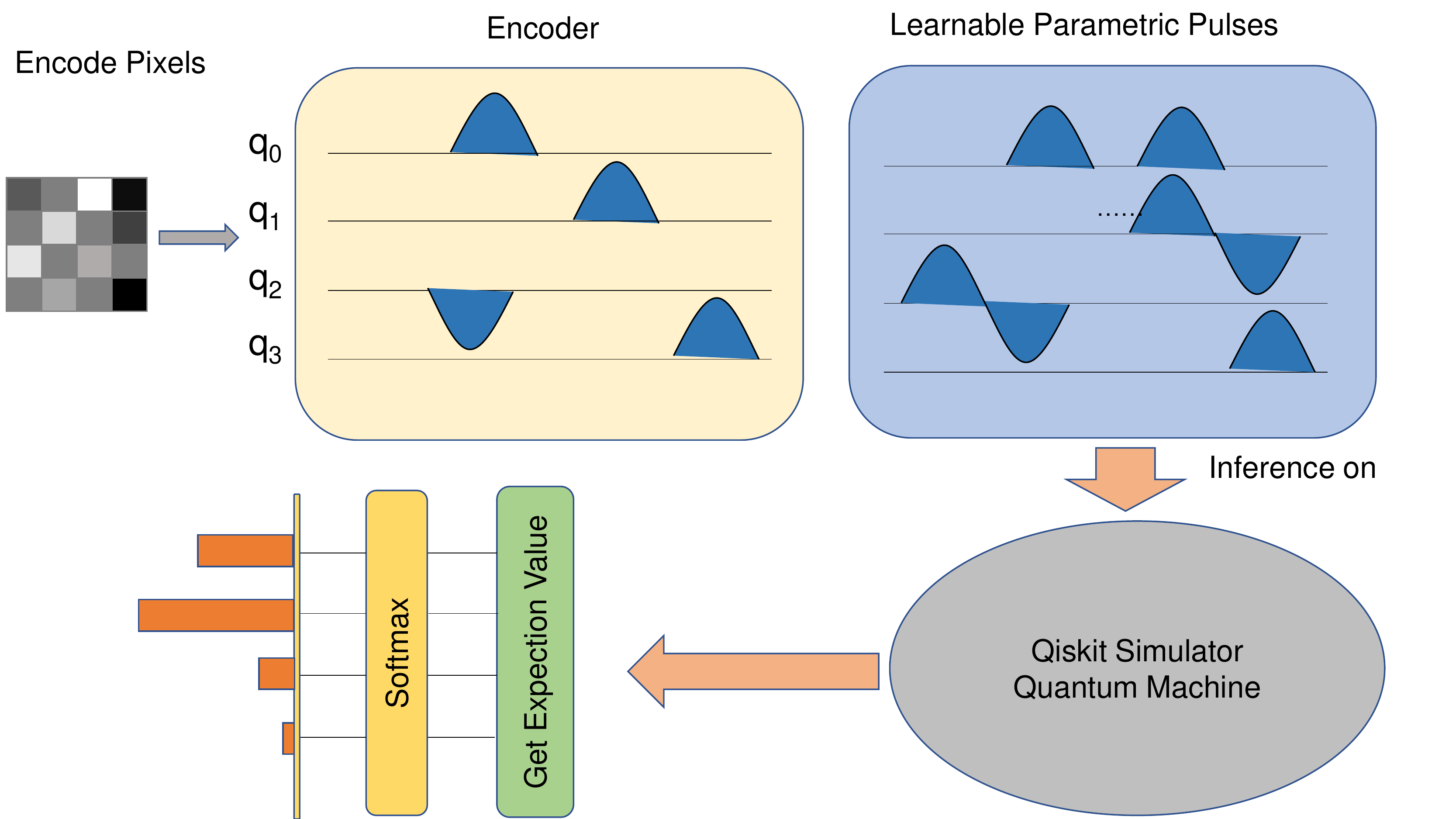}}
\caption{Conceptual illustration of VQP for QML tasks.}
\label{fig:VQP}
\end{figure}
\section{Background and Motivation}
\label{sec2}

\begin{figure}[t]
\centerline{\includegraphics[width=\linewidth]{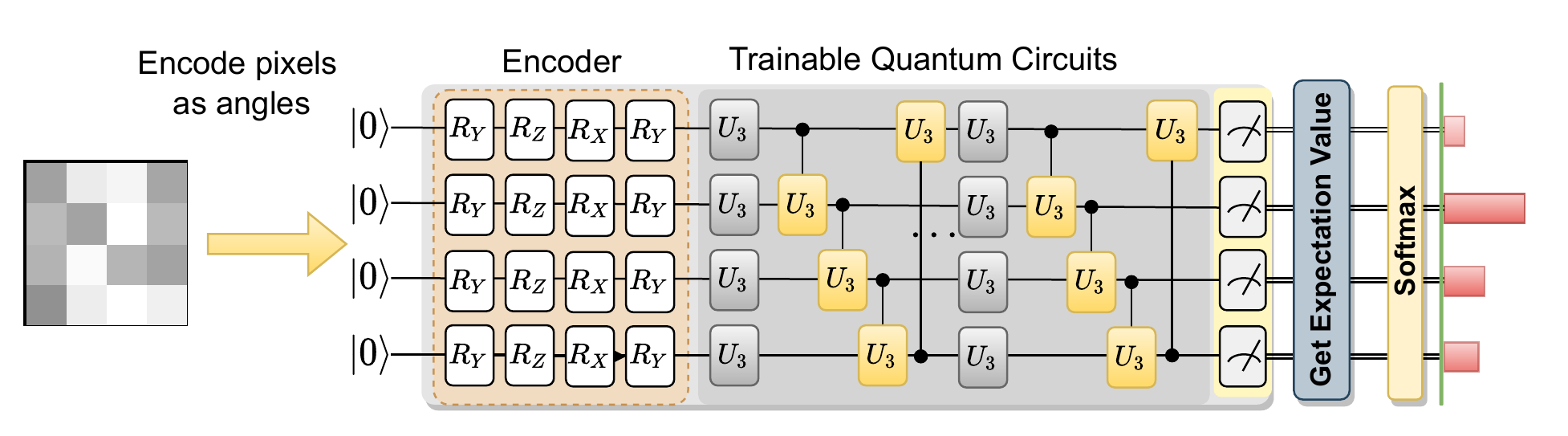}}
\caption{An example QNN that uses VQC for QML tasks.}
\label{fig:VQC}
\end{figure}
Our idea mainly comes from the usage of the VQC for the machine learning tasks. VQC enables quantum gates to be parameterized \cite{kandala2017hardware}, such as assigning a value to the angle of a rotation gate, thus making the quantum circuit trainable. The training parameter is actually the angle of the rotation gate, which can also be understood as phase shift training. More specifically, a variational quantum state $|\psi(x, \theta)\rangle = \Phi(x, \theta) |0......0\rangle$ is prepared by parameterizing the quantum circuit, where $x$ is the input data, $\theta$ is a set of free variables for training, and $\Phi()$ is parameterized quantum circuit. Usually, the VQC is trained by quantum-classical co-optimization to find the optimal set of parameters for the circuit. Quantum neural network (QNN) is a quantum machine learning model that uses VQC to encode features on the input data and then performs complex-valued linear transformations. For image classification tasks, it is necessary to first form a quantum encoding circuit by encoding pixels to several rotation gates, and then use VQC to compute and process the information passed to it. After the computation, we measure the qubits based on the Z-axis and get expectation values, and then calculate the corresponding probability, which is already a digital number because the measurement carries out the transformation from quantum state to classical data. VQC and QNN have great potential for applications in quantum machine learning (QML) \cite{biamonte2017quantum, ahsan2022quantum}, quantum simulation and optimization \cite{moll2018quantum, cheng2020accqoc}.
\subsection{Variational Quantum Pulse}

Inspired by the VQC, we propose the concept of the VQP. First of all, a general understanding is needed that the workflow of quantum computing at the software level is from the logic quantum circuits of high-level programming. Then the quantum circuits are mapped to the physical qubits after "transpiler". The quantum gates will further be "translated" into quantum pulses through a lookup table. These pulses are what quantum machine really corresponds to and processes with. In this paper, we define a VQP as a set of trainable quantum pulses that are parameterized, where a trainable pulse is defined as a pulse described by specific parameterization such as frequency, duration, amplitude, shape (e.g. Gaussian, square wave, etc.), etc.

\subsection{Variational Quantum Pulse Learning}
Also inspired by the fact that VQC can be trained, we believe that VQP has the potential to be trainable. However, the wide variety of parameters of VQP can make the search space for training too large, so that the whole training is not doable.
In this case, we narrow down the parameters of the attempted training to the amplitude item. As mentioned in the work \cite{wright2022deep}, a physical neural network can be constructed within a physical system in nature as long as parameters and an executable simulator exist, which also means that the parameters are trainable. And from the result of work \cite{meitei2021gate}, we can also make a guess that pulse should have the trainability.

Moreover, the training of amplitude is actually pulling and pushing the set of pulses in the pulse model without changing the shape of the pulse (the nature of quantum gates is mainly determined by the shape, e.g. the function of entanglement of CX gates is not affected by modifying amplitudes in a limited range). With the support of Qiskit's OpenPulse\cite{mckay2018qiskit}, We can take the amplitude both in the drive channel and control channel into VQP training to train the parameters that cannot be considered in the parameterized quantum circuit.  The control channels are device-dependent. The role of these channels is described by the Hamiltonian returned by the device, with enough details to allow its operation. Based on this property, making amplitudes in control channels learnable may give the ability for tolerant noise in each specific device. 
\begin{figure}[t]
\centerline{\includegraphics[width=\linewidth]{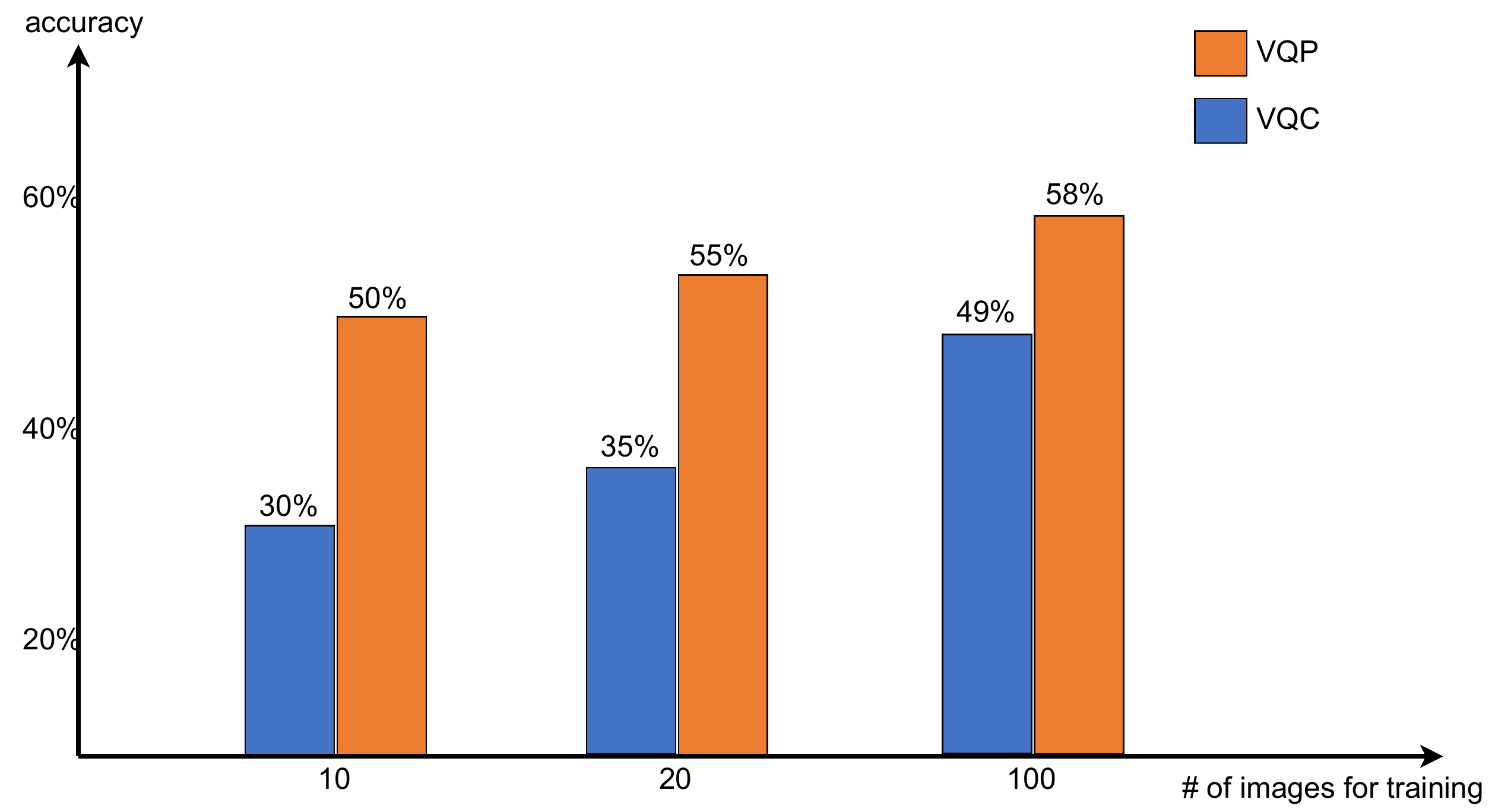}}
\caption{Comparison between VQC and VQP on a binary classification task from mini-MNIST datasets on qiskit pulse simulator with system model from ibmq\_quito.}
\label{fig:motivation}
\end{figure}

Furthermore, given the limited variety of quantum gates available in quantum circuits, we should not limit our vision to the existing quantum gates. The training of VQP can be viewed as a black-box optimization problem in which the quantum gates corresponding to the pulses in the optimized black box are unknown, but these unknown gates may have better results for our QML tasks. Figure \ref{fig:VQP} illustrates the concept to use VQP for 
classification tasks. The process is quite similar 
to QNN that uses VQC. We first use encoding circuit to encode the image pixels by putting them as angles in the rotation gate and transforming them to pulses format. Then we build the trainable pulses either can be transformed from a circuit (e.g. VQC) or put parameters in pulse builder, do the VQP training to process pulses, which can be in a pulse simulator or a real quantum machine. After that, we measure all the qubits and store the information in the acquire channel for calculating the expectation values by using Softmax. After that, we can obtain the probabilities based on expectation values.

We have set up a simple experiment with MNIST two-class classification to see if VQP learning can show some advantages over VQC learning. We use qiskit pulse simulator with system model from ibmq\_quito as the backend to process the quantum circuit and pulse schedule. Due to the limited resource on the quantum computer, we configured the experiments using 10, 20, 100 images for training and the same number of images for testing respectively. The baseline VQC is randomly generated by interleaving U3 and CU3 as shown in Figure \ref{fig:VQC}. The results for the three settings are shown in Figure \ref{fig:motivation}. It is obvious that in all the cases, VQP learning can achieve much better accuracy than VQC learning. 

Apprantely, the performance of VQP learning, like all other learning framerworks, heavily depends on how well the training is carried out. Yet a major and unique challenge for VQP learning is that back-propagation is not possible to be calculated from the backend of qiskit OpenPulse. Therefore, a non-gradient-based optimization framework to enable VQP learning is needed. This will be discussed in the following sections.
\begin{table}[]
\centering
\renewcommand*{\arraystretch}{1}
\setlength{\tabcolsep}{5.5pt}
\footnotesize
\begin{tabular}{cccc}
\toprule
\midrule
\multirow{3}{*}{Form of CX gate} & \multicolumn{3}{c}{Time Duration}   \\
                                 & Pulse simulator & Pulse simulator & Pulse simulator \\ 
                                 & (Quito) & (Belem) & (Jakarta)\\\midrule
CRX($\pi$) gate                     & 26832.0dt & 32016.0dt & 26832.0dt   \\ \midrule
CX gate                          & 25136.0dt & 27728.0dt & 25136.0dt   \\ \midrule
\bottomrule
\end{tabular}

\caption{Time duration for CX gate in different forms on pulse simulator with system models of ibmq\_quito, ibmq\_belem, and ibmq\_jakarta.}
\vspace{6pt}
\label{CXgate}
\vspace{-15pt}

\end{table}

\subsection{Advantages of VQP learning}
For quantum computers in the NISQ era, noise is a very big problem. VQC training architectures always need tens or even hundreds of gates to support learning. For VQP learning, pulse has more parameters, so the number of gates needed is greatly reduced. The reduction in the number of gates will directly lead to a decrease in latency, which means that the decoherence error can be reduced significantly. In this case, the process of VQP learning can be more robust to the environment of noise than VQC learning.

Moreover, the U gate is composed of some RZ gates as well as two fixed RX gates, RX($-\pi/2$) and RX($\pi/2$). This means that for circuit-level training of the U gate, the process of angle changes is only addressed on the Virtual Z and does not have a real impact on the physical amplitude. On the other hand, VQP learning, with amplitude as a parameter, actually changes the physical amplitudes of pulses.

Finally, VQP can also reduce the time duration for some special quantum gates. For example, in the gate level training,  a CX gate needs to be treated as a special CRX($\pi$) gate, the time duration of which is shown in Table \ref{CXgate} for pulse simulator with system model of ibmq\_quito, ibmq\_belem, and ibmq\_jakarta. While at the pulse level, we can train directly on the pulse of the CX gate, and the time duration is 
shorter on the same machine. 

\begin{figure}[t]
\centerline{\includegraphics[width=\linewidth]{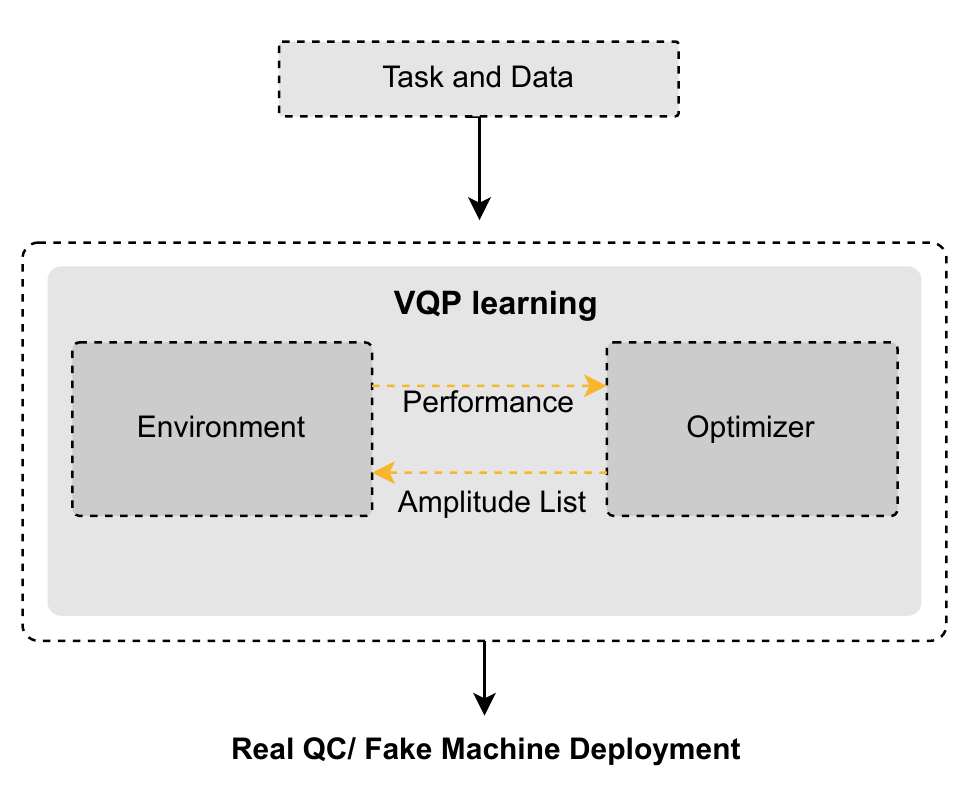}}
\caption{Overview of the VQP learning framework.}
\label{fig:workflo}
\end{figure}

\section{VQP Learning Framework}
\subsection{Overview}
\label{sec3}

In this section, we present an overview of the VQP learning framework. As shown in Figure \ref{fig:workflo}, we use optimizer continues communicate with the environment. We first get a parameterized quantum circuit as a baseline and transform it into a pulse schedule. Then we put this set of pulses into the pulse optimization framework and give the amplitude list of pulses and the corresponding accuracy list of pulses to the optimizer as the initial parameters. After finding the optimized amplitude list with the lowest error rate, we need to reconstruct the pulse based on this amplitude list. The reconstructed pulses are finally deployed and executed on the pulse simulator in the qiskit test mock or on a real quantum device. Subsection B below describes the circuit to pulse transform process. Subsection C proposes the optimization framework for VQP learning. Subsection D illustrates the process of pulse reconstruction based on the optimized amplitudes.

\subsection{Quantum Circuit \& Pulse Transform}
The baseline for this work is a VQC that can also be used for QML tasks. An example is shown in Figure \ref{fig:VQC}. Since there is no existing method for pulse encoding, we choose to continue to use the circuit encoding approach to encode classical data into quantum states. Then we transform the encoded circuit part, which already contains the input data and tasks, into a pulse schedule. This process can be achieved in qiskit's OpenPulse. After successfully encoding the data, we transform the trainable circuit part into pulse schedule by the same method in order to ensure the fairness of the comparison in the experiments. After the transformation, we get two parts of the pulse schedule: encoding pulses and variational quantum pulses, which can also be considered as trainable pulses. Then we need to fix these encoding pulses, and only variational quantum pulses are considered in the following training process.
\subsection{Pulse Optimization Framework}
\begin{figure}[t]
\centerline{\includegraphics[width=\linewidth]{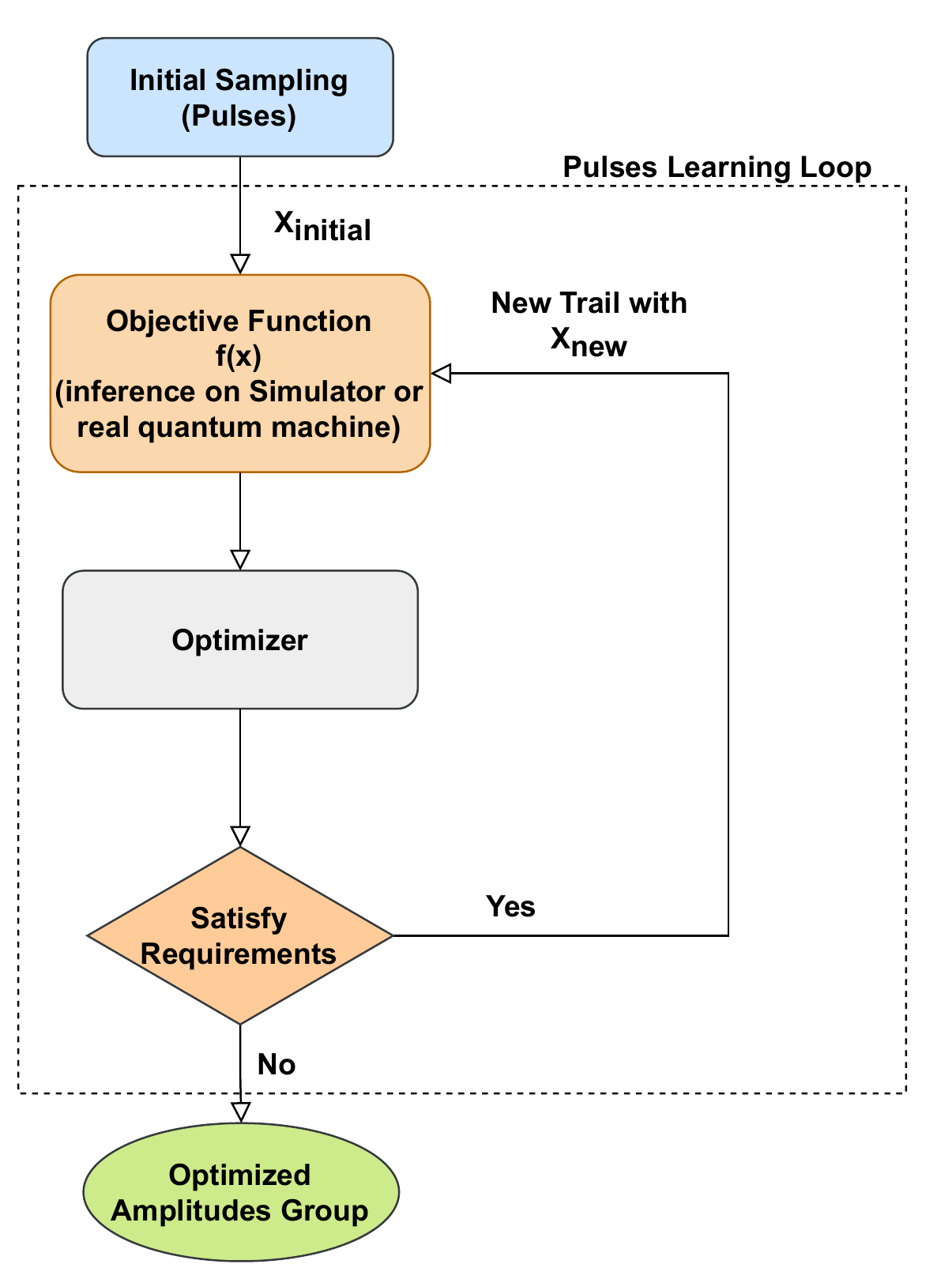}}
\caption{The schematic of pulse optimization framework.}
\label{fig:BO}
\end{figure}

A wide range of algorithms can be possible candidates for this framework, such as RL evolutionary algorithms, genetic algorithms~\cite{kennedy1995particle,Ingber1993SA}, Bayesian optimization~\cite{gao2022bayesopt,gao2019bayesopt2,shahriari2016bayesopt}, etc. For this optimization framework, we extract the amplitudes of the pulses that are obtained as described in the previous subsection. At this point, the amplitudes are complex numbers, which means that the pulse optimization framework cannot handle it. But from the amplitude, we can associate the magnitudes and angles of these complex numbers. Then we combine the computed magnitudes and angles into a single array, which can be trained and optimized by the pulse optimization framework. It is worth noting that we need to revert this array back to the form of the complex numbers in the objective function to obtain the amplitude and then reconstruct pulses, because this objective function should be evaluated on a qiskit pulse simulator or a real quantum machine. Doing so allows our pulse optimization framework to tolerate random noise, which we believe can gain benefits in NISQ era quantum devices.

Here, we use Bayesian optimization (BO)~\cite{gao2022bayesopt,gao2019bayesopt2,shahriari2016bayesopt} to explain the optimization framework: combining the property of pulses, we propose a pulse BO framework. As shown in Figure \ref{fig:BO}, we first take the amplitude list extracted from pulse as the initial input for the BO. After the quantum machine executes, we can get an error rate, denoted as the initial objective values, also part of input for the BO. At the same time, we need to set the search space for BO. For an executable pulse, its norm needs to be no larger than one.
We build a Gaussian Process (GP) regression model with RBF kernel based on the initial amplitude list, yielding a surrogate model for the real simulation. The model calculates the mean and variance of the function values at each point. With the GP model accessible, we constructs and optimize an acquisition function, which is used to decide the location to sample points. We choose the lower confidence bound (LCB) as the acquistion function in this work with the aim of finding the minimum error rate. After finding the minimum error rate, we can know the current searched best list, denormalize these data, and then recalculate the magnitude and angle of the current searched best list into amplitude, after which we get the optimized amplitude list.
\begin{algorithm}[t]
\caption{Variational Pulse BO Learning}\label{alg:one}
\KwData{$\rho$, $\chi$, $M$, $D$}
// $\rho$ is the amplitude list, $\chi$ is the search bound, $D$ consists of $x_i$ and $y_i$, $M$ is the Gaussian Process Regression model. \\
$D \gets $InitPulses$(\rho, \chi)$\;
\For{$i \gets |D|$ to $N\_total$}{ // iterative optimization
    $p(y|x,D) \gets FitModel(M, D)$\;
    // Acquisition function actively searches for the next optimized amplitudes.\\
    $x_i$ $\gets$ $argmin_{x \in \chi}S(x, p(y|x,D))$\;
    // Calculate corresponding error rate by processing in quantum machine.\\
    $y_i$ $\gets$ $f(x_i)$\;
    $D$ $\gets$ $D \bigcup (x_i, y_i)$\;
}
\end{algorithm}

Algorithm \ref{alg:one} presents the simplified pseudo code of the proposed variational pulse BO learning framework. $\rho$ is the amplitude list from initial pulses for optimization, $\chi$ is the search bound, constraint the norm of amplitude of pulses need to be less or equal to one, $D$ is consists of $x_i$ and $y_i$, $x_i$ is the hyper-parameter and is amplitude list during optimization in this case, and $y_i$ is the error rate that results by executing $x_i$ by quantum machines, $M$ is the Gaussian Process Regression model. By using $\rho$ inference on quantum device backend, we can get $(x_i, y_i)$ pairs, $y_i$ is the error rate in this case. And we set $N_{total}$ as the iteration times for optimization, we set the Gaussian process regression model, set the kernel method with normalized $y_i$, and fit the normalized $x_i$ to the Gaussian form. Then, $S$ is proposed as an acquisition function to select parameters by constraints, aiming at minimize $y$. Finally we compute the updated $y_i$ with updated $x_i$.

\subsection{Pulse Reconstruction}
After obtaining the optimized amplitude list, we design a waveform reconstruction function. This function first stores the optimized amplitude in the modified list. Then it extracts the amplitude values from the initial pulses and overwrites the initial amplitude values one by one by iteratively storing the optimized amplitude in the modified list. The other parameters in the initial pulses are kept unchanged, so as to build new pulses with only the amplitude change.

\section{Experiments}
\label{sec4}
\subsection{Experiments Setup}
\textbf{Dataset:} We evaluate the proposed method using the binary classification task, which is the same as that used in \cite{jiang2021co}.
Specifically, it is a binary classification.
In generating the dataset, we will associate 1 out of 2 classes to two input values ($x$ and $y$).
For example, we associate class 0 to inputs of x = 0.2 and y = 0.6; and class 1 to x = 0.8 and y = 0.8.
On top of the created dataset, we will divide them to to the train set and test set. And we also use the MNIST dataset, we do center-cropped images to 24 × 24, and then down-sample to 4×4 for two-class. We also setup a method to make sure the number of images we get from different classes keeping same.
% conduct experiment on binary classification tasks from \cite{jiang2021co}, due to the limited real QC and noise simulator resources, we use limited data of ``train''and ``test'' split as train and test set. 
% We stream the inputs to VQP, and obtain the output $p$.
% If $p\geq 0.5$, it is classified as class 0; otherwise, it is class 1.
% For this binary classification task, we mark the output probability of 
% class 0 as $p_0$, if $p_0 \geq 0.5$, our input can be recognized as class 0, otherwise, will be recognized as class 1. 

\textbf{VQP Framework and Baseline:} 
We apply the gate based QNN as the baseline, which is implemented by TorchQuantum, an existing library for quantum machine learning. 
Figure \ref{fig:VQC} shows the detailed QNN structure of the baseline.
% For the encoding circuit part, we use four RY, four RX, four RZ, and four RY gates acting on the four qubits respectively. These sixteen rotation gates encode the data into quantum states. The encoding circuit is transformed into pulse schedule form as the pulse encoder part of the experiment. For the VQC part of the QNN, we use U3 and CU3 layers interleaved as in Figure \ref{fig:VQC}, and the total gates used in the VQC part is nine. 
In VQP framework, we obtain the initial pulse by converting the baseline QNN, and then, we will learn the parameters in the pulse to iteratively
% And we also convert it into pulse form as the VQP part. 
% The variational quantum pulses are put into the framework of VQP learning, and the optimization process of 
tuning pulses 
% is performed 
on the pulse simulator/real quantum devices. 
The number of iterations is set as 30. With a trained VQP, we can stream the test set to get the accuracy. 
In the evaluation, we apply both gradient method and the same BO framework for VQC training.
The optimization objective is to minimize the error rate (i.e., $1 - accuracy$). Limited by the property of optimizer, for training parameter in high dimensional space, the optimization results is unstable, so we run five seeds for every benchmark and calculate the average as the result we reporting in this paper.

% In the optimization process, we aim at minimizing the error rate, so in the training framework, we get the error rate by $1 - accuracy$. For the VQC training, we mostly do the gradient-based training with 15 epochs, but we also consider the BO framework for VQC training with 15 iterations.

\begin{figure}[t]
\centerline{\includegraphics[width=\linewidth]{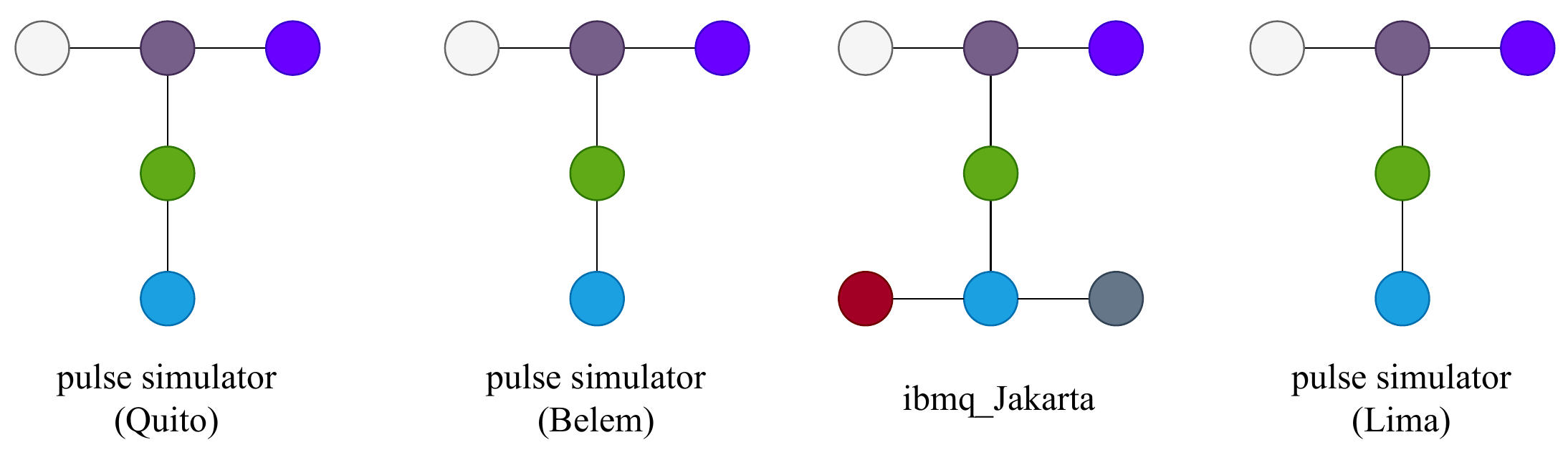}}
\caption{The coupling map of all quantum devices we use.}
\label{fig:coupling}
\end{figure}

\textbf{Measurement of VQP Learning: }After run the pulse schedule in quantum machine, we can obtain the quantum state $\phi$ from computational qubit $|q\rangle$. 
By setting the shots as $x$, say $x=256$, we can obtain the probabilities of results to be state $|0\rangle$ (corresponding to output of +1) and $|1\rangle$ (corresponding to output of -1).
% The quantum machine can return us the counts for $|0\rangle$ and $|1\rangle$ since $q \in {0, 1}$. We set output is +1 when we get $|0\rangle$ and output is -1 when we get $|1\rangle$.Thus, we can compute expectation value by repeating the experiments for $x$ shots, and during this work, we set the shots as 256, and 
For example if we get 156 $|0\rangle$ and 100 $|1\rangle$, then we can calculate the probability of -1 as $(156 * 1 + 100 * (-1))/256 = 0.21875$.

\textbf{Quantum Hardware and Backend Setting: }We use the pulse simulator and quantum computer provided by IBM Quantum.
First, we employ pulse simulators as pulse simulator (Quito), pulse model (Belem), and pulse simulator (Lima), these simulators can be imported from the qiskit test mock, which are the simulators that pull the real data from the real machine. These settings are due to pulse simulator do not consider the physical error from qubits, but it will occur algorithmic error during Hamiltonian estimating. And as we want to keep circuit learning and pulse learning in a fair environment, we run both in pulse simulator, pulse level we do training on pulse and processed in pulse simulator in every optimization iteration, and for circuit level, we do training on circuit and transform the optimized circuit to pulse then processed in pulse simulator in every optimization iteration.
Second, we also conduct the evaluation on ibmq\_jakarta, a real quantum machine that can process pulse schedule. 
% The optimization level is set to two for all experiments, and the measure level is also set to two. 
Kindly note that noise is considered in both pulse simulator (algorithmic error) and real machine (physical error from qubits), indicating we evaluate the proposed 
% So, we can test the performance of the proposed 
method in a noisy environment.
\begin{table}[t]
\centering
\renewcommand*{\arraystretch}{1}
\setlength{\tabcolsep}{5pt}
\footnotesize
\begin{tabular}{ccccccc}
\toprule
\hline
\multirow{3}{*}{Task Data} & \multicolumn{3}{c}{Accuracy}  \\ 
                        & Pulse simulator     & Pulse simulator   & Pulse simulator  \\
                        &(Quito)  & (Belem) & (Lima) \\ \midrule

Initial 20              & 0.45         & 0.45          & 0.5         \\
\textbf{+ VQP learning}     & \textbf{0.65} & \textbf{0.6} & \textbf{0.55}\\ \midrule
Initial 100             &    0.51      &    0.5      &       0.5    \\ 
\textbf{+ VQP learning} & \textbf{0.57} & \textbf{0.63} & \textbf{0.61}              \\ \midrule
Initial MNIST 20             &   0.5       &    0.5      &   0.45        \\ 
\textbf{+ VQP learning} & \textbf{0.55} & \textbf{0.66} & \textbf{0.6}              \\ \midrule
Initial MNIST 100             &    0.5      &    0.5     &  0.51       \\ 
\textbf{+ VQP learning} & \textbf{0.57} & \textbf{0.61} & \textbf{0.61}              \\ \midrule
        \bottomrule
\end{tabular}

\caption{Binary classification task results on different dataset size and run on pulse simulators with system models from ibmq\_quito, ibmq\_belem, ibmq\_lima, separately.}
\vspace{6pt}
\label{result1}
\vspace{-15pt}
\end{table}

\begin{table}[t]
\centering
\renewcommand*{\arraystretch}{1}
\setlength{\tabcolsep}{6.5pt}
\footnotesize
\begin{tabular}{ccccl}
\toprule
\midrule
\multirow{2}{*}{Model} & \multicolumn{2}{c}{Accuracy}  \\ 
                      & Pulse simulator (Belem) & ibmq\_jakarta  \\ \midrule
VQC learning 20         &  0.57   &     0.58                            \\ 
\textbf{VQP learning 20} & \textbf{0.6} & \textbf{0.69}                  \\ \midrule
VQC learning  100        & 0.61      & 0.59                      \\ 
\textbf{VQP learning 100} & \textbf{0.63} & \textbf{0.64}                    \\ \midrule
VQC learning  MNIST 20        & 0.6      & 0.56                      \\ 
\textbf{VQP learning MNIST 20} & \textbf{0.66} & \textbf{0.62}                    \\ \midrule
VQC learning  MNIST 100        & 0.57      & 0.62                      \\ 
\textbf{VQP learning MNIST 100} & \textbf{0.61} & \textbf{0.71}                    \\ \midrule
\bottomrule
\end{tabular}
\caption{Classification tasks results by VQP learning VS VQC learning with same BO framework on pulse simulator with system model from ibmq\_Belem and ibmq\_jakarta.}
\vspace{6pt}
\label{result2}
\vspace{-15pt}
\end{table}
\begin{table}[t]
\centering
\renewcommand*{\arraystretch}{1}
\setlength{\tabcolsep}{25.5pt}
\footnotesize
\begin{tabular}{ccc}
\toprule
\midrule
Model                          & \# of Gates             & Accuracy \\ \midrule

VQC\_base   & 9         &   0.62       \\ \midrule

\textbf{VQP}         & \textbf{9}        & \textbf{0.71}          \\ \midrule
VQC* & 12 & 0.68 \\ \midrule
\bottomrule
\end{tabular}
\caption{Comparison the MNIST two-class classification task with 100 images and result on ibmq\_jakarta between VQC\_base, VQP, VQC* with different number of gates and all by same BO framework.}
\label{result3}
\end{table}
\begin{table}[t]
\centering
\renewcommand*{\arraystretch}{1}
\setlength{\tabcolsep}{20pt}
\footnotesize
\begin{tabular}{ccc}
\toprule
\midrule
Model        & \# of Gates & Accuracy      \\ \midrule
\textbf{VQP} & \textbf{9}  & \textbf{0.71} \\ \midrule
VQC with gradient   & 9           & 0.73          \\ \midrule
VQC* with gradient  & 12          & 0.77          \\ \midrule
\bottomrule
\end{tabular}
\caption{Put VQC with different gates in gradient based method and VQP in the BO framework, test on ibmq\_jakarta and get results.}
\vspace{6pt}
\label{WithBO}
\vspace{-15pt}
\end{table}
% \usepackage{multirow}
% \usepackage{graphicx}
\begin{table}[t]
\centering
\renewcommand*{\arraystretch}{1}
\setlength{\tabcolsep}{5pt}
\footnotesize
\begin{tabular}{cccc}
\toprule
\midrule
\multirow{2}{*}{Model}   & \multirow{2}{*}{\# of Gate} & \multicolumn{2}{c}{Time Duration}       \\  
                         &                             & ibmq\_jakarta      & Pulse simulator (Belem)          \\ \midrule
\textbf{VQP}             & \textbf{9}                  & \textbf{40816.0dt} & \textbf{45168.0dt} \\ \midrule
VQC*                     & 12                          & 58896.0dt          & 58768.0dt          \\ \midrule
\textbf{VQP\_transpiled} & \textbf{11}                 & \textbf{32368.0dt} & \textbf{32816.0dt} \\ \midrule
VQC*\_transpiled         & 17                          & 53008.0dt          & 46192.0dt          \\ \midrule
\bottomrule
\end{tabular}
\caption{Time duration for VQP, VQC*, VQP\_transpiled, and VQC*\_transpiled based on ibmq\_jakarta and pulse simulator with system model of Belem.}
\vspace{10pt}
\label{models}
\vspace{-15pt}
\end{table}

\begin{table}[t]
\centering
\renewcommand*{\arraystretch}{1}
\setlength{\tabcolsep}{2.6pt}
\footnotesize
\begin{tabular}{lccc}
\toprule
\midrule
\multicolumn{1}{r}{\small \underline{Use system model of $\rightarrow$}} &  \multirow{2}{*}{Pulse simulator (Lima)}  & \multirow{2}{*}{Pulse simulator (Quito)}  \\
        {\small \underline{Inference on $\downarrow$}} & & \\
        \midrule

        Pulse simulator (Lima) & \cellcolor{blue!20} \textbf{0.65} & \cellcolor{red!20} 0.4  \\
        Pulse simulator (Quito) & \cellcolor{red!20} 0.35 &  \cellcolor{blue!20} \textbf{0.55} \\
\midrule
\bottomrule
\end{tabular}
\caption{Run and test accuracy result on different models.}
\vspace{6pt}
\label{dependence}
\vspace{-15pt}
\end{table}
\subsection{Main Results}

\textbf{VQP Learning Result: }
% We experiment on binary classification tasks running on three quantum machines using the framework of VQP learning to demonstrate the effectiveness of VQP learning.
We take the initial pulses inference on quantum machine's result as an initial result, and we set the different size of tasks: (1) 20 'train' data and 20 'test' data, which is described as 'Initial 20'; (2)  'Initial 100' for 100 'train' data and 100 'test' data; (3) 'Initial MNIST 20' for 20 'train' image and 20 'test' image in MNIST two-class; (4) 'Initial MNIST 100' for 100 'train' image and 100 'test' image in MNIST two-class.

% in Table \ref{result1}. Meanwhile, 'Initial 20' and 'Initial 100' are "20 images for 'train', 20 images for 'test'" and "100 images for 'train', 100 images for 'test'".

Table \ref{result1} report the experimental results. VQP learning can improve accuracy for all settings. 
% If we remove the 10 images case which can be taken as less trustable results,
Specifically, compared with 'Initial 100', VQP learning can get an average improvement of 10\% accuracy.
The improvement of VQP is 12\% over 'Initial MNIST 20'.
% on binary classification task with 20 images. It can also get an average improvement of 43.33\% accuracy on a task of 100 images. 

% These three noise simulators are all using real data including noise from real quantum machine, which means our VQP learning obtain 100\% accuracy under noise environment. 

\textbf{Comparison between VQP and VQC: }We also compared the results of VQC and VQP on the same task. VQC uses the nine-gate learnable circuit.
% before the transformation to pulse. 
For a fair comparison, we performed 30 iteration same BO on two benchmarks, pulse simulator with system model of ibmq\_belem, and real quantum machine ibmq\_Jakarta, executing 256 shots each.
% This is set to correspond to the 15 iterations of VQP learning, the same benchmark, and the number of shots executed to ensure the same overhead.

As can be seen from the Table \ref{result2}, the results of VQP learning achieved 60\% and 69\% accuracy for the 20-data. VQC learning, on the other hand, obtained 57\% and 58\% accuracy in pulse simulator (Belem) and ibmq\_jakarta. VQP learning show better performance in the 8 groups tasks, and achieve up to 71\% accuracy on the MNIST two-class 100 image classification task, meanwhile, VQC learning also gain the best performance on this task but with 62\% accuracy.
% results were obtained in FakeBelem and ibmq\_jakarta, respectively.
An observation is that VQP learning obtain up to 94\% accuracy under noise environment in real quantum machine, but sometimes only push performance to gain a little bit beneifts. This is because of non-gradient based optimizer sometimes cannot handle the hyperparameters in high dimensional space. Thus, a differentiable simulator is highly demand for the pulse learning. However, from another aspects that we can say 94\% is at least a second optimal solution that can be achieve by VQP learning. 
\begin{figure*}[t]
\centerline{\includegraphics[width=\linewidth]{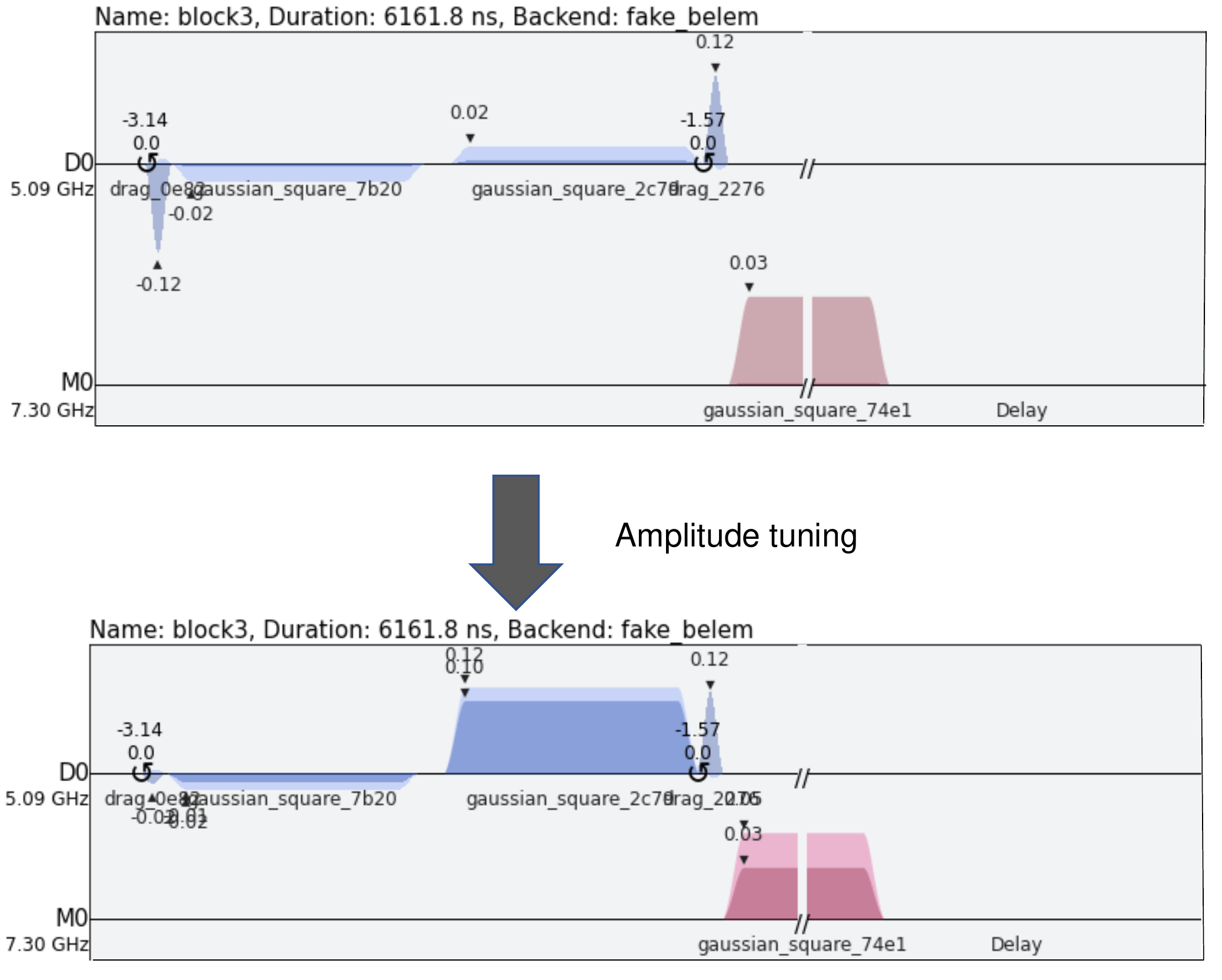}}
\caption{Pulse visualization of CX gate before and after amplitude tuning in pulse simulator (Belem).}
\label{fig:vis}
\end{figure*}
\begin{figure}[t]
\centerline{\includegraphics[width=\linewidth]{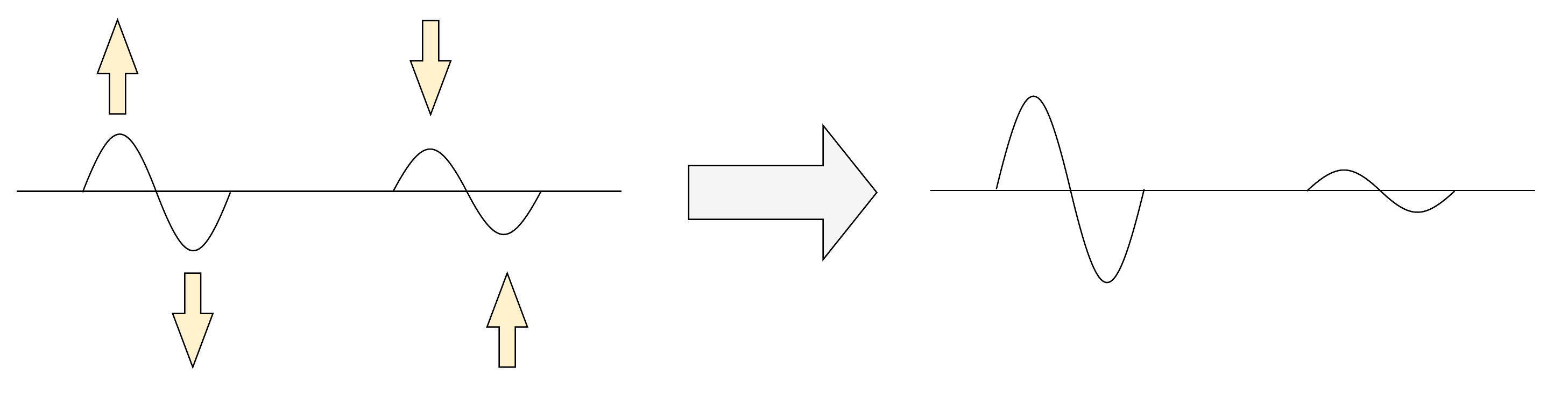}}
\caption{Schematic of varying the amplitude of pulse.}
\label{fig:vary}
\end{figure}

Overall, from the above results, VQP learning optimized for parameter amplitudes at pulse level can show advantage over VQC learning trained on parameter angles at circuit level under the same conditions. 
% This demonstrates the potential of our proposed new paradigm, and we believe that VQP learning has even greater potential. 
% Our current VQP is based on an unoptimized VQC architecture. If it is based on an optimized VQC architecture such as QuantumNAS \cite{wang2021quantumnas}, we expect that VQP learning can achieve better performance.

\textbf{Further Experiment on VQP and VQC: }To verify the conjecture of VQP learning advantage in the Section \ref{sec2}. We conduct another set of experiment to compare VQP and VQC learning.
% Here, we still ensured fairness on the overhead as well as the fairness of the environment. 
As can be seen from the Table \ref{models}, VQC* with 12 gates achieves 68\% accuracy on MNIST two-class classification, while VQC\_base with nine gates achieves only 62\% accuracy. 
This indicates that the training results get improved when the number of gates of VQC increases (i.e., more parameters); however, VQP learning with only 9 gates still outperform VQC learning with 12 gates. 
This observation illustrate
% can well prove the previous speculation 
that VQP learning has more trainable parameters than the gate based VQC for better learning. 

In the Table \ref{WithBO}, we report the results of VQC learning in gradient based method on ibmq\_jakarta.
It is obviously that training VQC in gradient based method can gain the benefits. And now VQC learning with 9 gates already has the better performance than VQP leanring in terms of accuracy.
% still achieve lower accuracy than . 
% The VQC* using BO framework can push the accuracy to 92\%, and still lower than the result we obtain from VQP learning, We believe these results can further support the advantage of the VQP learning. 

\textbf{Device Dependence of VQP: }VQP is device-dependent. As shown in Table \ref{dependence}, if we train on pulse simulator (Lima) and test on pulse simulator (Lima), we can obtain 65\% accuracy, but if we train on pulse simulator (Lima) and test on pulse simulator (Quito), the accuracy is only 35\%, which shows that device-specific is important. This is because the pulse for the same gate can vary from model to model. And from the work \cite{proctor2022measuring}, we think in the quantum machine the device dependence of VQP can also result in the noise is structured-dependent.

\section{Discussion}
\label{sec5}
We further discuss the purpose of extracting pulse amplitudes from the pulse schedules. Here, we provide the visualization on the change of variational quantum pulses in the process of optimization. Based on Equation \ref{eq1}, the Hamiltonian of the control pulses will be updated accordingly in the training process.
Moreover, the latency of 12 gates is necessarily larger than that of 9 gates.
On ibmq\_jakarta and pulse simulator (Belem), we tested the time duration of 9 gates VQP against 12 gates VQC*, which have similar accuracy for both using BO framework.
% during our experiments on ibmq\_jakarta and FakeBelem, respectively. 
% The transpile is set with the same initial and the optimization level is two. 
From the table, we can see that the time duration of VQP for 9 gates is much shorter than that of VQC* for 12 gates in all different cases. 
From this result, we can intuitively see the advantage of VQP learning over VQC learning in terms of latency, which means that the robustness of VQP learning on noise not only comes from the optimization framework but also the property that its decoherence error is smaller.

\begin{equation}
\begin{aligned}
H = \sum_{i=0}^{1} (U_i(t)+D_i(t)) \sigma_i^{X} +
\sum_{i=0}^{1} 2\pi \nu_i (1-\sigma_i^{Z})/2 \\+
\omega_B a_B a^{\dagger}_B +
\sum_{i=0}^{1} g_i \sigma_i^{X} (a_B + a_B^{\dagger})
\label{eq1}
\end{aligned}
\end{equation}

\textbf{Vary of Amplitudes of Pulses: }
As shown in Figure \ref{fig:vary}, the process of VQP learning is equivalent to pulling or pushing initial pulses. In the process of iterative optimization, multiple attempts are made to vary amplitudes until the error rate is minimized. 
Figure \ref{fig:vis} show the visualization of the pulse of CX gate on pulse simulator (Belem). The plot on the left is the pulse of the CX gate before amplitdue tuning, while the plot on the right is the new pulse after amplitdue tuning. 
This shows how the proposed framework to directly tune the physical amplitudes.

\textbf{Analytical Understanding of Amplitude Tuning: }In the process of VQP learning, we can analyze the physical quantities affected by amplitude tuning according to Equation \ref{eq1}. This equation describes the drive Hamiltonian, where $D_i(t)$ is mixed by the signal on drive channel for qubit $i$ and local oscillator (LO) at frequency corresponding to the signal. $U_i(t)$ is mixed by the signal on control channel for qubit $i$ and some combinations of qubit LO's that specified by device. $\sigma_X, \sigma_Y$ and $\sigma_Z$ are Pauli operators. $\nu_i$ is the estimated frequency of qubits in the qubit i, $g_i$ is the coupling strength between qubits, $\omega_B$ is the frequency of buses, $a_B$ and $a^{\dagger}_B$ are the ladder operator for buses. The vectors actually effected by amplitudes tuning are $D_i(t)$ and $U_i(t)$. $D_i(t)$ and $U_i(t)$ can be obtained by the Equation \ref{eq2}: 

\begin{equation}
\begin{aligned}
D_i(t) = Re (d_i(t)e^{iw_{d_i}t})\\
U_i(t) = Re [u_i(t)e^{i(w_{d_i} - w_{d_j}) t})]
\label{eq2}
\end{aligned}
\end{equation}
where $d_i(t)$ and $u_i(t)$ are the signal of qubit $i$ on drive channel and control channel, respectively. Amplitude is the intensity of signal, which means when we do amplitudes tuning, we change the intensity of signal, so that affect on the vary of $d_i(t)$ and $u_i(t)$. It is known from Equation \ref{eq2} that the changes in $d_i(t)$ and $u_i(t)$ affect the $D_i(t)$ and $U_i(t)$ in Equation \ref{eq1}.

\section{Conclusion and Prospective}
\label{sec6}
For QNN, its potential advantage over classical neural networks is that the search space of the unitary matrix can be increased with the number of qubits, such that the neural network can learn more and performs better. For VQP learning, the pulse has more parameters than the circuit, which means pulse learning may obtain better expressibility and entangling capability. The focus of this work is to propose a novel paradigm to use VQP for quantum learning. We demonstrate the advantages of VQP learning over VQC learning for ML tasks. Also, the reduced decoherence error due to the reduced latency from the small demand for gate number in VQP learning is really important for quantum machines in the NISQ era, since noise is one of the major problems in the NISQ era.

The potential of VQP is huge. It can be more flexible to tune and process the parameters inside the circuit. As we discussed in the experimental section, it currently achieves results only based on an unoptimized VQC architecture, and we expect to see better results if we apply it on an optimized circuit-level architecture.
The current VQP learning has pitfalls, one being the limited resources of the real quantum machine and the other being that we do not have an effective simulator for pulses. The simulator that can support pulses in the OpenPulse interface provided by Qiskit is slow to execute, which makes it difficult to experiment with larger and more complex tasks at the moment. As well, this pulse simulator is not differentiable, which also leaves us at the moment to use non gradient based optimizer in VQP learning, which tend to be more stochastic in parametric tasks in high dimensional spaces. Therefore an efficient and differentiable pulse simulator is urgently needed.

In addition, training and optimization methods for VQP are worth investigating, and we plan to implement a more efficient optimization and training framework in the future. Gradient-based machine learning for VQP learning may still be possible with advances in the supporting platforms.

\section*{Acknowledgment}
We thanks Thomas Alexander for patient guided on qiskit OpenPulse, also thanks and Dr. Xiangliang Zhang for valuable discussion about the optimization framework. We acknowledge the use of IBM Quantum services for this work.

\printbibliography
\end{document}